\documentclass[twocolumn]{aa} 
\usepackage{epsfig}

\usepackage{natbib}

\usepackage{amssymb}

\usepackage{lineno}

\begin{document}
\title{Neptune's Atmospheric Composition from AKARI Infrared Spectroscopy}

\author{L.N. Fletcher\inst{1,2} \and P. Drossart\inst{3}
          \and M. Burgdorf\inst{4} \and G.S. Orton\inst{2}  \and T. Encrenaz\inst{3}}

\institute{Atmospheric, Oceanic and Planetary Physics, Clarendon Laboratory, University of Oxford, Parks Road, Oxford, OX1 3PU\\
	\email{fletcher@atm.ox.ac.uk}\\
\and
Jet Propulsion Laboratory, California Institute of Technology, 4800 Oak Grove Drive, Pasadena, CA, 91109, USA
\and
LESIA, Observatoire de Paris, UPMC, Universite Paris-Diderot, 5 place Jules Janssen 92195 Meudon, France
\and      
SOFIA Science Center, Deutsches SOFIA Institut, NASA Ames Research Center, Mail Stop 211-3, Moffett Field, CA 94035, USA.
}

   \date{}

 
  \abstract
   {}
   {Disk-averaged infrared spectra of Neptune between 1.8 and 13 $\mu$m, obtained by the AKARI Infrared Camera (IRC) in May 2007, have been analysed to (a) determine the globally-averaged stratospheric temperature structure; (b) derive the abundances of stratospheric hydrocarbons; and (c) detect fluorescent emission from CO at 4.7 $\mu$m.}
   {Mid-infrared spectra (SG1 and SG2 channels of AKARI/IRC), with spectral resolutions of 47 and 34 respectively, were modelled using a line-by-line radiative transfer code to determine the temperature structure between 1-1000 $\mu$bar and the abundances of CH$_4$, CH$_3$D and higher-order hydrocarbons.  A full non-LTE radiative model was then used to determine the best fitting CO profile to reproduce the fluorescent emission observed at 4.7 $\mu$m in the NG channel (with a spectral resolution of 135).}
   {The globally-averaged stratospheric temperature structure is quasi-isothermal between 1-1000 $\mu$bar, which suggests little variation in global stratospheric conditions since studies by the Infrared Space Observatory a decade earlier.  The derived CH$_4$ mole fraction of $(9.0\pm3.0)\times10^{-4}$ at 50 mbar, decreasing to $(0.9\pm0.3)\times10^{-4}$ at 1 $\mu$bar, is larger than that expected if the tropopause at 56 K acts as an efficient cold trap, but consistent with the hypothesis that CH$_4$ leaking through the warm south polar tropopause (62-66 K) is globally redistributed by stratospheric motion.   The ratio of D/H in CH$_4$ of $3.0\pm1.0\times10^{-4}$ supports the conclusion that Neptune is enriched in deuterium relative to the other giant planets.  We determine a mole fraction of ethane of $(8.5\pm2.1)\times10^{-7}$ at 0.3 mbar, consistent with previous studies, and a mole fraction of ethylene of $5.0_{-2.1}^{+1.8}\times10^{-7}$ at 2.8 $\mu$bar.  An emission peak at 4.7 $\mu$m is interpreted as a fluorescent emission of CO, and requires a vertical distribution with both external and internal sources of CO.  Finally, comparisons to previous L-band studies indicate significant variability of Neptune's flux densities in the 3.5-4.1 $\mu$m range, related to changes in solar energy deposition.}
   {}

\titlerunning{AKARI Spectroscopy of Neptune}
   \maketitle
%

\section{Introduction}

Disk-integrated spectra of Neptune between 1.8 and 13 microns were recorded on May 13, 2007, using both the prism and the grism modes of the Infrared Camera \citep[IRC,][]{07onaka} on board ISAS/JAXA's AKARI infrared astronomy satellite \citep{07murakami}, which launched on February 21, 2006 (UT).  The broad wavelength-coverage provided by AKARI permits simultaneous observation of near-IR reflectance of sunlight and mid-IR thermal emission.  These data are used to constrain models of Neptune's atmospheric structure, composition and aerosols in the troposphere and stratosphere.  

Interior models indicate that Neptune's atmospheric envelope of H$_2$ and He represents only a thin shell compared to the radius of the planet \citep[e.g.][]{99guillot}, but the collision-induced continuum of H$_2$-He is the dominant source of opacity in the mid-IR spectrum (6-13 $\mu$m).  In addition to continuum emission, the 6-13 $\mu$m spectrum exhibits substantial structure due to the chemistry of stratospheric methane (CH$_4$), CH$_3$D and higher-order hydrocarbons (Fig. \ref{spectrum}).  Neptune's bulk C/H ratio is known to be enriched over the solar abundance of \citet{07grevesse} by $54\pm15$ \citep[from the measurements of][]{95baines}.  This is consistent with formation models \citep[see, e.g.][]{04hersant, 06owen}, which predict that Neptune received a larger proportion of heavy elements (relative to the H$_2$-He envelopes) compared to the other giant planets.  AKARI spectroscopy of both CH$_4$ and the higher-order hydrocarbons will be used to determine Neptune's stratospheric temperature and composition. The cold atmospheric temperatures yield an extremely low-power thermal emission spectrum which requires high sensitivity to measure accurately, and AKARI results will be compared to previous derivations of Neptune's thermal structure from the Voyager radio science investigation \citep[RSS,][]{90lindal_nep}, infrared spectrometer \citep[IRIS,][]{98conrath} and the Infrared Space Observatory \citep[ISO,][]{99feuchtgruber, 03fouchet, 03burgdorf_nep}.

Previous studies have shown that the stratospheric abundance of CH$_4$ is larger than the saturated abundance at the tropopause \citep{94baines}, which should serve as an efficient `cold trap' for CH$_4$, restricting it to the troposphere.  Vigorous vertical advection of both gaseous CH$_4$ and CH$_4$ ice (which subsequently sublimates) was invoked to explain the stratospheric abundance.  Such rapid convection is consistent with observations of Neptune's strong internal heat flux,  the disequilibrium of para-hydrogen \citep{98conrath}, the high levels of tropospheric CO determined from the sub-mm \citep{07hesman} and with the visible observations of thick, high clouds casting shadows on deeper levels in Voyager 2 images.   However, convective cloud activity is typically restricted to Neptune's mid-latitudes and is time-variable \citep[e.g.][]{01sromovsky}, and it remains unclear whether the strength of the upwelling is sufficient to maintain the stratospheric CH$_4$, which is continually destroyed by photolysis.  As an alternative, ground-based mid-IR imaging \citep{07orton_nep} demonstrated the presence of a hot south polar region, warm enough to permit significant CH$_4$ escape into the stratosphere without the need for strong convection.  Ultimately, it is likely that both mechanisms contribute, but AKARI derivations of the global CH$_4$ abundance will be used to reassess the requirement for strong convection of methane-laden air into the stratosphere.

Once in the stratosphere, CH$_4$ is photolysed to form the higher hydrocarbons.  Acetylene (C$_2$H$_2$ at 13.7 $\mu$m) and ethane (C$_2$H$_6$ at 12.2 $\mu$m) were first detected in ground-based spectroscopy \citep{87orton} and confirmed by Voyager/IRIS spectra \citep{91bezard}.  Ethylene (C$_2$H$_4$ at 10.5 $\mu$m) was detected in ISO spectra by \citet{99schulz}, and both methylacetylene (C$_3$H$_4$ at 15.8 $\mu$m) and diacetylene (C$_4$H$_2$ at 15.9 $\mu$m) were recently detected in Spitzer/IRS spectra \citep{08meadows}.  These hydrocarbons may diffuse downwards and freeze out as haze layers, and sunlight absorption on these hydrocarbon hazes may be partially responsible for the warm temperatures of the lower stratosphere.

The near-IR region of the AKARI spectrum (1.8-5.0 $\mu$m) is dominated by the reflection of sunlight from cloud and haze layers, with broad absorptions due to atmospheric CH$_4$.  Clouds of CH$_4$ are expected to condense at the 1-2 bar level, overlying much deeper clouds of NH$_4$SH and H$_2$O, but only two of the uppermost layers have been observed:  a thin cloud near 1.5 bar and an optically thick one with cloud tops near 3.8 bar \citep[from ground-based H$_2$ quadrupole lines and CH$_4$ reflectance spectroscopy][]{94baines, 95baines}.  The near-IR spectrum will be used to determine the vertical distribution of CO from a fluorescent peak detected at 4.7 $\mu$m.  

The AKARI/IRC data reduction methods for both the near-IR and mid-IR spectra  are described in Section \ref{spectra}.  In Section \ref{midir} we use the mid-IR spectra to determine the global stratospheric temperatures, the abundances of CH$_4$ and CH$_3$D and estimates for the hydrocarbon mole fractions.  The temperature structure is used in Section \ref{nearir} to model the fluorescent emission of CO.  The implications of each of these new compositional measurements will be discussed in Section \ref{discuss}.

\section{Observations}  
\label{spectra}

\begin{figure}[htb]
\begin{center}
\epsfig{file=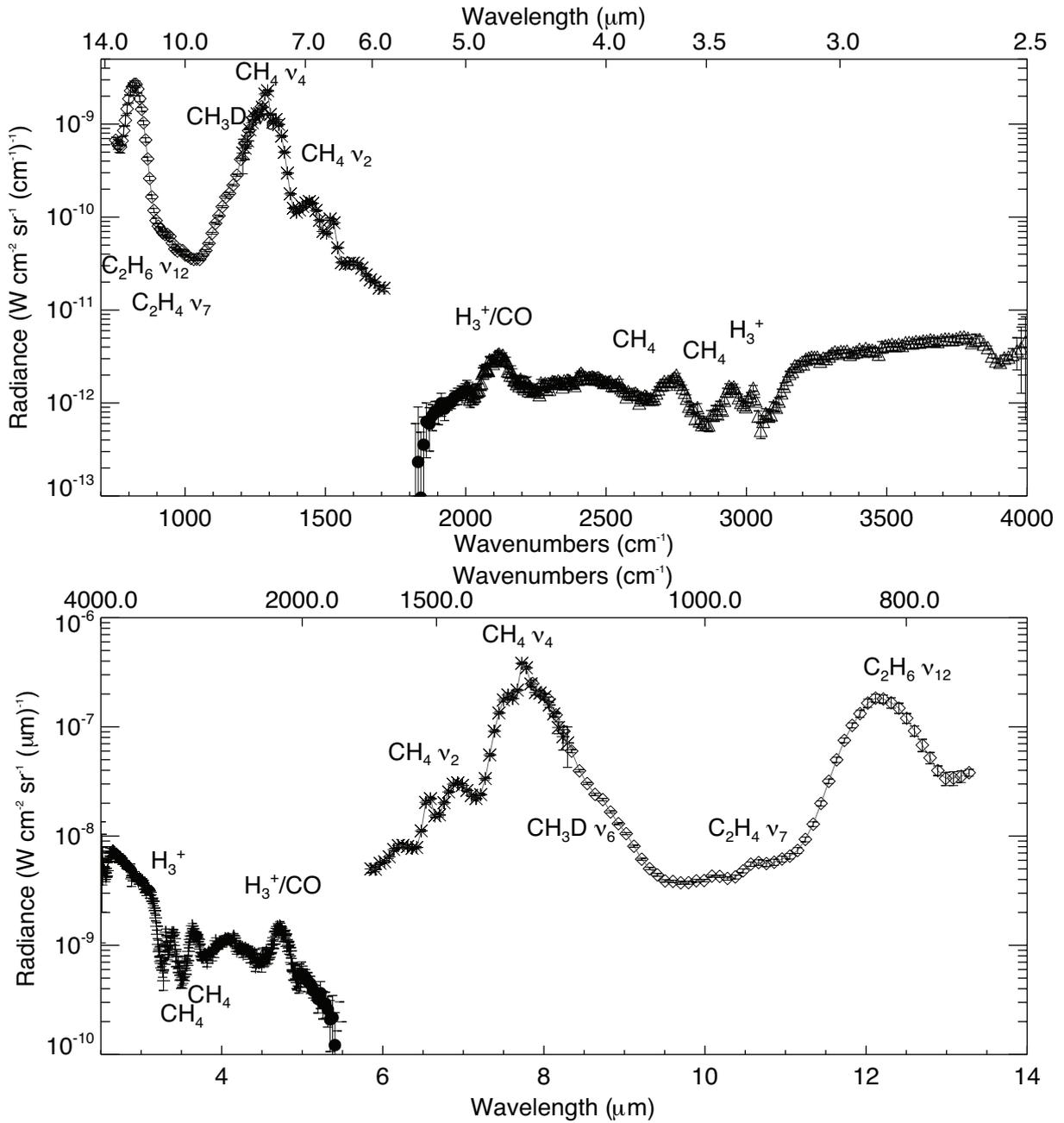,width=16cm}
\caption{AKARI/IRC Spectrum of Neptune from 13 May 2007.  The upper panel plots the radiance per wavenumber, which emphasizes the near-IR portion of the spectrum from the NP (circles) and NG (triangles) modes.  The lower panel plots the radiance per wavelength, which emphasizes the mid-IR spectrum from the SG1 (crosses) and SG2 (diamonds) modules.  SG1 data between 5.5-5.8 $\mu$m and NP data shortward of 5.0 $\mu$m are omitted for clarity.}
\label{spectrum}
\end{center}
\end{figure}

Fig. \ref{spectrum} shows the near and mid-IR spectra of Neptune on May 13, 2007, with many of the emission and absorption features described above.  The emission features of CH$_4$ and C$_2$H$_6$ are particularly prominent in the mid-IR spectra, with CH$_4$ absorptions visible in the reflected sunlight component of the near-IR.  In addition, we label peaks which may include some H$_3^+$ emission (although these cannot be quantified properly in the low-resolution AKARI spectra) and a fluorescent emission due to CO at 4.7 $\mu$m (see Section \ref{nearir}).

These disk-averaged spectra were obtained in four modules (Table \ref{obs}):  SG1 (5.5-8.3 $\mu$m) and SG2 (7.4-13 $\mu$m) channels in the mid-IR, and prism (NP, 1.7-5.5 $\mu$m)  and grism modes (NG, 2.5-5.0 $\mu$m) in the near-IR.  The spectral resolving power was 135 in the interval 2.5 - 5 $\mu$m and between 20 and 50 at other wavelengths.  The two grisms on the mid-IR channel observe simultaneously, whereas either the NP or NG modes must be selected on the near-IR channel.  The lower-resolution near-IR prism spectrum is not shown in Fig. \ref{spectrum} except for the 5.0-5.5 $\mu$m region, but the measured flux was found to be consistent between the NG and NP observations except between 4.1 and 4.7 $\mu$m, where the values from the prism are systematically some 20\% higher than those obtained with the grism. This may simply be due to an error in the flux calibration of the slitless spectrum or it may imply temporal or spatial variations in Neptune's atmosphere (Section \ref{meteo}). Long-term variations by a factor of two in the L~band have been observed already in the past \citep{03feuchtgruber}.

Raw data were reduced using version 2.0 of the IRC spectral reduction pipeline, which includes masking software to avoid saturation in the mid-IR at bright hydrocarbon emission features.  Each pointing consisted of eight exposures (nine for the grism in the near IR), and any sub-frames with systematic noise patterns were removed from the analysis.  As Triton was included in the field of view at a distance of only 6 arcsec in the cross-dispersion direction, we had to subtract its contribution to the near-infrared spectra.  For the prism and grism modes, Triton's contribution was obtained from off-source spectra, extracted at the same distance from the moon as Neptune but on the other side, and then subtracted.  The contribution of Triton to the prism mode was found to be much larger than to the grism mode, because in the slitless spectrum Triton stood ``at longer wavelengths'' relative to Neptune in the dispersion direction.  As a consequence, the subtraction of Triton from the NG spectrum would have only served to increase the spectral noise, so was omitted in this reduction.    NG shows good agreement with ISO at 2.7 $\mu$m \citep[based on][]{00encrenaz_esa}.  

The contribution due to Triton was deemed negligible in the mid-IR, so it has not been subtracted from the spectra.  Spectra were extracted from a 3-pixel wide column for SG1 and 5 pixel column for SG2.  The short-wavelength end of SG1 (shortward of 5.8 $\mu$m) demonstrated a rising flux which was deemed spurious (because the signal-to-noise ratio was rather low in this region), hence the 5.5-5.8 $\mu$m region is omitted from Fig. \ref{spectrum}.  As the nominal range of the NP spectrum ends at 5.2 $\mu$m, there is a discontinuity in the spectrum between five and six microns.

\subsection{Field of View}

AKARI does not spatially-resolve Neptune's disc, which subtended $9.45\times10^{-11}$ steradians at the time of the observations.  With the exception of the NG grism mode (where we used a slit 1~arcmin wide), all spectra were obtained in a slitless mode (10'x10') so that Neptune was fully included in the FOV.  The fact that Neptune is an extended source (2.3 arcsec compared to a 1.5-arcsec pixel size for the NIR camera) was taken into account during the reduction.  For the near-IR spectra of Neptune we used 3 pixels along the cross-dispersion axis and checked that there was negligible signal outside the extraction rectangle.  3 pixels are the default (this is the FWHM of the inflight point spread function in the near-IR), and using a wider column would have further contaminated the spectrum with flux from Triton.  In the mid-IR, Neptune's diameter is even more negligible compared with the instrumental PSF than in the near-IR.

\subsection{Wavelength Calibration}

The pipeline determines the location of various sources and uses a linear relation between pixel location and wavelength.  The slitless nature of the AKARI/IRC, in addition to observing an extended source and the problem of saturation in longer exposures which can mask the precise location of a source, can lead to difficulties when determining the wavelength calibration.  Furthermore, as spectra can only be extracted from the array on an integer-pixel basis, this can introduce systematic pixel errors as large as $\pm0.03$ $\mu$m.  The positional uncertainty for wavelength calibration in the NP spectrum is also of the order of a single pixel, leading to a wavelength calibration offset of approximately 0.05 $\mu$m. 

\subsection{Instrument Function}

Information on the spectral lineshape was not available at the time of writing, so we use the assumption of a Gaussian with the required spectral resolution,  $R=\lambda/\Delta \lambda$, where $\Delta \lambda$ is the smallest resolving element, which should be a constant in each module to a first approximation.   A comparison between spectra computed with a constant $R$ and a constant $\Delta \lambda$ indicated that this approximation was valid.  The specified resolutions are given in Table \ref{obs}.


\section{Analysis of Mid-IR Thermal Emission:  5.8-13 $\mu$m}
\label{midir}

Spectra from the AKARI/IRC SG1 and SG2 mid-IR channels were analysed using an updated version of the line-by-line radiative transfer algorithm, RADTRANS \citep{97irwin}.  The disk-averaged transmission was calculated via an exponential integral method \citep{89goody}.  Because the longest wavelength accessible to AKARI is 13 $\mu$m, we are unable to use the smooth collision-induced H$_2$ and He continuum longwards of 15 $\mu$m to perform a full derivation of the tropospheric temperature structure beneath 0.1 bar.  However, AKARI spectra are sensitive to the stratospheric temperatures in the 1-1000 $\mu$bar region via the emissions of the CH$_4$ and the higher-order hydrocarbons.  The following sections describe the determination of stratospheric temperature and composition from the disk-averaged spectra.

\subsection{Reference Neptune Atmosphere}

As AKARI spectra offer no constraint on the atmospheric temperatures beneath the tropopause, we follow \citet{08meadows} and use the Neptune $T(p)$ structure from \citet{05moses_jup} as a starting point for fitting temperatures above the 1-mbar level.  The initial profile was formed from a combination of results from Voyager radio science investigation in the troposphere \citep[RSS,][]{90lindal_nep}, with updates in the stratosphere from the Infrared Space Observatory \citep{98bezard, 99feuchtgruber, 03fouchet, 03burgdorf_nep}  and from 17-24 $\mu$m ground-based spectra \citep{87orton, 90orton}. The resulting stratospheric profile became isothermal at 168 K above the 0.1-mbar level, but was forced to $250\pm50$ K at 0.1 $\mu$bar \citep[][and references therein]{98bezard} for consistency with stellar occultation data.  The profile was regridded onto a pressure scale between 10 bar and 0.1 $\mu$bar in 100 intervals equally spaced in $\log(p)$.  The gravitational acceleration at 1 bar is 11.2 ms$^{-2}$ when we take a global average (from equator to pole, where $g$ is strongest), and the forward model accounts for the variation of Neptune's gravity with latitude. The H$_2$ mole fraction was set to 0.83 with a helium mole fraction of 0.15 \citep{93conrath_nep}.  

The reference temperature profile used in this analysis had been used as input for a one-dimensional model of Neptune's stratospheric photochemistry \citep[Model A from][]{05moses_jup} which provided the hydrocarbon abundance profiles for scaling analysis in this data.  The deep CH$_4$ abundance was set to 2.2\% following \citet{94baines} who determined this value using a combination of the Voyager/RSS temperature profile and ground-based measurements of H$_2$ quadrupole lines.  The vertical CH$_4$ reference profile follows a saturated vapour pressure profile with 100\% humidity in the tropopause region, but then increased back to a mole fraction of 2.2\% when the stratosphere became warm enough. Above this level, the CH$_4$ abundance decreased according to the photochemical modeling of \citet{05moses_jup}.   Vertical hydrocarbon distributions were included following the analysis of \citet{05moses_jup}, and scaled to reproduce the observed spectra.  We hold $^{12}$C/$^{13}$C constant at the terrestrial value of 89.  Because AKARI spectra are insensitive to the S(0) and S(1) lines in the far-IR, the ratio of ortho-H$_2$ to para-H$_2$ was set to thermochemical equilibrium at every level of the atmosphere.  We used this reference atmosphere to calculate the contribution function (the product of the rate of change of the transmission with altitude with the black body function), which shows the level of emission for each of the spectral regions covered by the AKARI spectrum in the 0.3-3000 $\mu$bar range (Fig. \ref{cf}).  The sources of spectroscopic line data are the same as those used in Jupiter and Saturn analyses \citep{09fletcher_ph3, 09fletcher_ch4}.


\subsection{Methane Emission at 7.7 $\mu$m}

The spectral resolution of the AKARI mid-IR data does not permit characterisation of the stratospheric temperature independently of composition.  Instead, we must rely on a trade-off analysis between two strongly anti-correlated variables - the stratospheric temperature and the methane mole fraction, which both determine the strength of the emission of the $\nu_4$ CH$_4$ band at 1304 cm$^{-1}$.  Such a correlation leads to large error estimates on the derived quantities.  Furthermore, the vertical variation of CH$_4$ and the other hydrocarbons must be fixed to the models of \citet{05moses_jup}, because we have insufficient spectral resolution to determine the vertical variability of these species.  However, as the following analysis shows, the independent measurements of AKARI confirm the validity of the profiles used by \citet{05moses_jup}.

\begin{figure}[htb]
\begin{center}
\epsfig{file=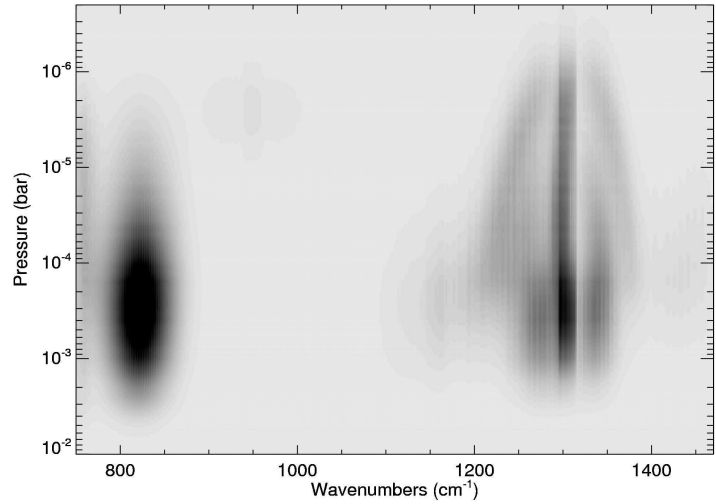,width=10cm}
\caption{Contribution function for the SG1 and SG2 spectral range, calculated using the reference atmosphere of \citet{05moses_jup}.  The darkest colours indicate the peaks of the contribution function.   Note in particular the multi-lobed nature of the CH$_4$ contribution function at 7.7 $\mu$m, used to derive the stratospheric temperature structure. }
\label{cf}
\end{center}
\end{figure}

\begin{figure}[htb]
\begin{center}
\epsfig{file=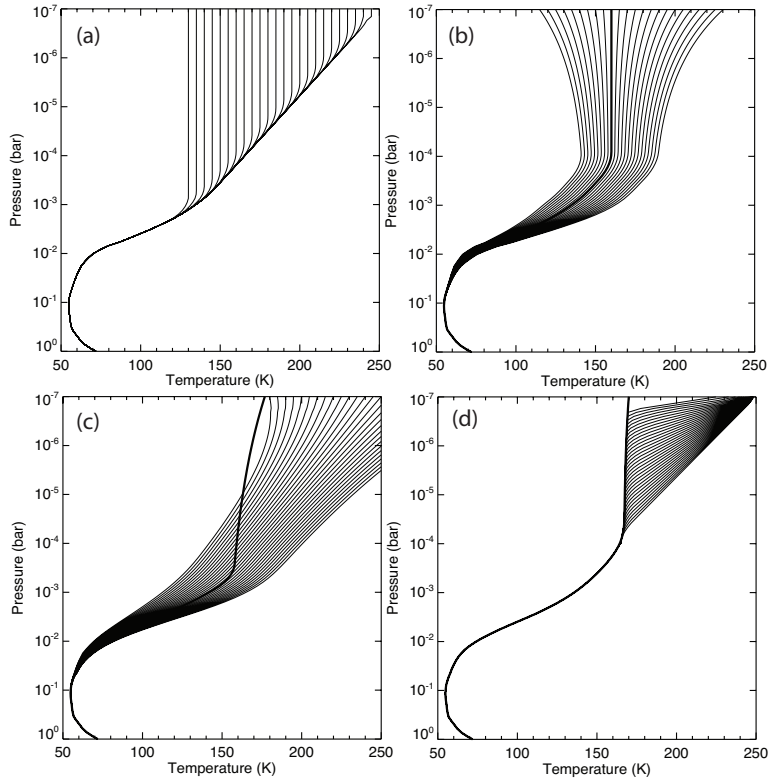,width=10cm}
\caption{Family of models (a-d described in the main text) for the derivation of Neptune's quasi-isothermal stratospheric temperatures.  The $T(p)$ profile for pressures greater than 1 mbar was set to the profile used by \citet{05moses_jup}, and the $T(p)$ in the 1-1000 $\mu$bar region was varied simultaneously with the stratospheric CH$_4$ abundance to fit the 7.7 $\mu$m emission band.}
\label{Tperturb}
\end{center}
\end{figure}

Our approach to the derivation of $T(p)$ was a family of single-variable parameterisations, each one building upon the last (Fig. \ref{Tperturb}) in a bottom-up direction (i.e. starting with the 1-mbar level where the temperature is best known from previous determinations, and working our way to 1 $\mu$bar and above).   This approach was guided by the multi-lobed nature of the contribution function in Fig. \ref{cf}, which shows sensitivity throughout the 1-1000 $\mu$bar range.  We attempt to preserve the smooth nature of $T(p)$ inferred from ISO spectra \citep{98bezard, 03fouchet, 03burgdorf_nep}, but we caution that the presence of real oscillations in the temperature profile (due to stratospheric wave activity, for example) would not be detected via this simple parameterisation. At the same time, we scaled the CH$_4$ abundance above the tropopause and minimised the fit to the data, $\chi^2$, over the 7.1-8.3 $\mu$m (1200-1400 cm$^{-1}$) spectral range.  The four temperature profile parameterisations were as follows;
\begin{enumerate}
\item Isotherm:  Starting from a smooth extrapolation between the 1 mbar level (approximately 136 K) and 0.1 $\mu$bar \citep[approximately 250 K from ][]{93yelle}, we generate a fully isothermal stratospheric temperature at a range of test values between 130 and 250 K  (Fig. \ref{Tperturb}(a)).  The best-fitting solution required 160-165 K and a CH$_4$ mole fraction of $1.0-3.5\times10^{-3}$ at 50 mbar.
\item Lapse Rate: Starting from the best-fitting isothermal structure, we add small positive and negative $dT/dp$ gradients as shown in Fig. \ref{Tperturb}(b), and smoothing the vertical structure so that there is some small perturbation at pressures greater than 1 mbar.  The best-fit solution required a warmer temperature profile increasing from 166 K at 100 $\mu$bar to 170 K at 1 $\mu$bar, and a smaller CH$_4$ abundance of $(1.0\pm0.5)\times10^{-3}$ at 50 mbar. 
\item Smoothing the Profile:  To further test the validity of a quasi-isothermal temperature structure, we use the family of $T(p)$ profiles in Fig. \ref{Tperturb}(c) to interpolate smoothly over the 1-1000 $\mu$bar region.  However, we found that the spectral fits were considerably worsened by the high temperatures at the $\mu$bar level, so these models were omitted from the subsequent analysis.
\item Extrapolation to 0.1 $\mu$bar: The model from step two reproduce the spectrum everywhere except for the centre of the CH$_4$ Q-branch, which probes the highest altitudes (Fig. \ref{cf}).  Preserving the quasi-isotherm in the 10-1000 $\mu$bar region, the profiles in Fig. \ref{Tperturb}(d) were used to test our sensitivity to the high 0.1 $\mu$bar temperatures (approximately 250 K) determined from stellar occultations.   Satisfactory solutions were found for a transition from the quasi-isotherm to the warm temperatures at altitudes between 0.1-5.0 $\mu$bar, with a CH$_4$ mole fraction of $(6.0\pm3.0)\times10^{-4}$ at 50 mbar.
\end{enumerate}

\begin{figure}[htb]
\begin{center}
\epsfig{file=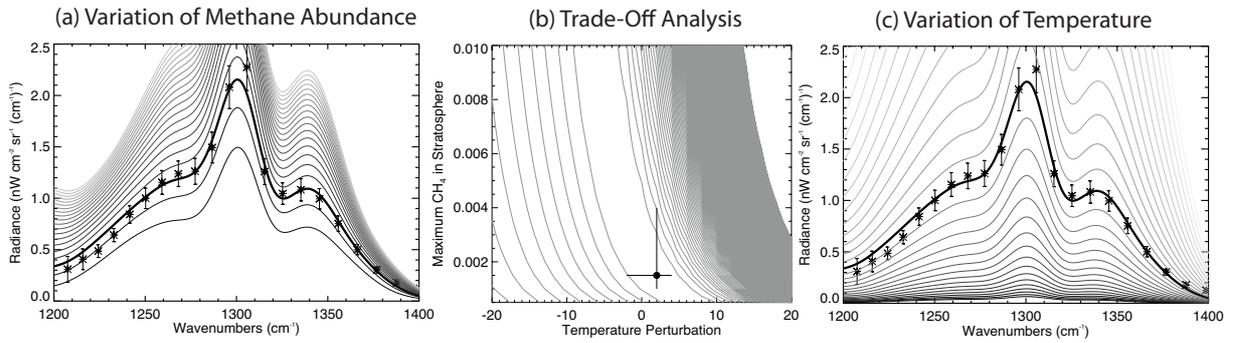,width=16cm}
\caption{Demonstrating the trade off analysis for fitting temperatures and methane with the family of models in Fig. \ref{Tperturb}.  Panels (a) and (c) show the competing effects of varying CH$_4$ and $T(p)$, and panel (b) shows how the best fitting model and error range was estimated from a 2-dimensional $\chi^2$ analysis.} 
\label{Tch4}
\end{center}
\end{figure}

\begin{figure}[htb]
\begin{center}
\epsfig{file=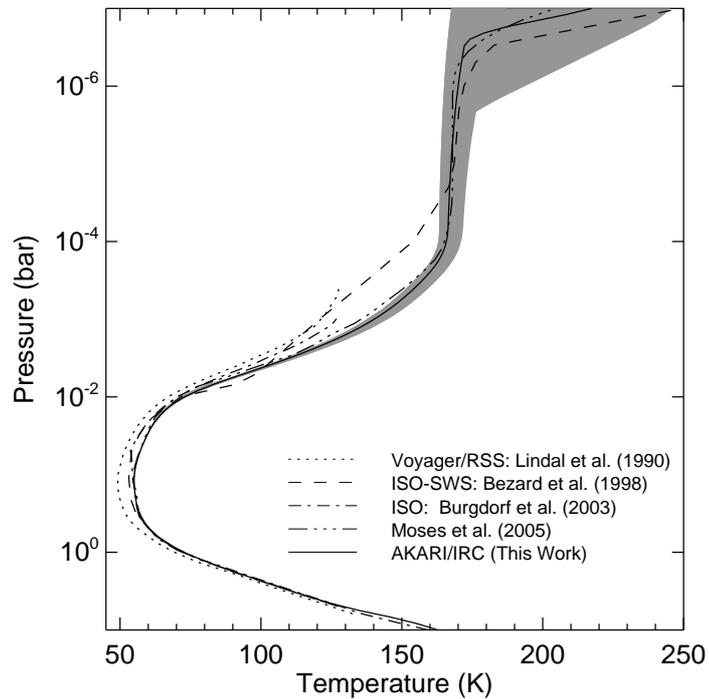,width=10cm}
\caption{Best-fitting temperature profile (solid line) compared to the results of a selection of Neptune analyses.  The grey shading indicated the range of possible solutions in the stratosphere given that the {\bf CH$_4$} abundance is highly anticorrelated with 
temperature (see main text).  Nevertheless, the tradeoff analysis in Fig. \ref{Tch4} provides stratospheric temperatures consistent with previous analyses from ISO and Voyager.}
\label{temp}
\end{center}
\end{figure}

An example of the $\chi^2$ analysis is shown in Fig. \ref{Tch4}, which shows a two dimensional $\chi^2$ surface in the centre, the spectral effect of varying the CH$_4$ mole fraction on the left and the temperature profile on the right.  The two correlated parameters can be marginally separated because of the different morphologies of their effects on the spectrum.  The best-fitting temperature structure and the error range (taking the anticorrelation with CH$_4$ into account) are shown in Fig. \ref{temp}, compared to $T(p)$ determinations from previous studies.  The warm and cool extrema of the error range were used in the error analysis for composition in the following sections.  The best-fitting temperature profile provides an optimum CH$_4$ mole fraction of $(9.0\pm3.0)\times10^{-4}$ at 50 mbar, decreasing to $(0.9\pm0.3)\times10^{-4}$ at 1 $\mu$bar.  However, using the cool and warm extrema of the temperature profile, the CH$_4$ mole fraction could vary between $6.0-18.0\times10^{-4}$.    The temperature and CH$_4$ determinations will be discussed in Section \ref{discuss}.

\subsection{CH$_3$D at 8.6 $\mu$m}

\begin{figure}[htb]
\begin{center}
\epsfig{file=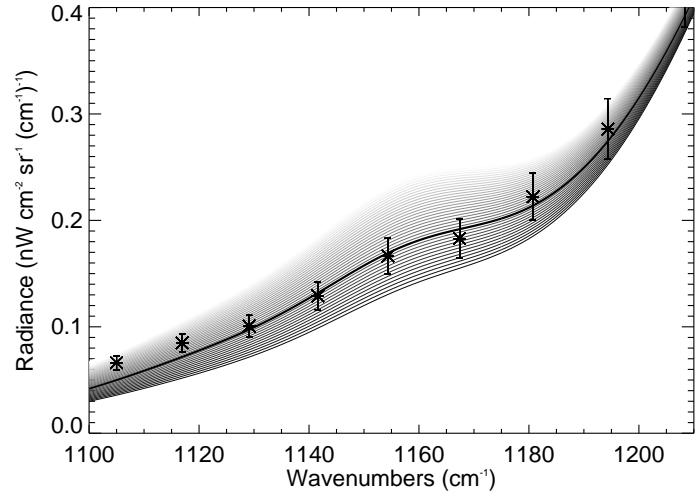,width=10cm}
\caption{Varying the abundance of CH$_3$D has a small effect on the spectrum at 1160 cm$^{-1}$.  Although the spectral lines of CH$_3$D are not observed, we can broadly fit the shape of the smooth spectrum to estimate the D/H ratio in stratospheric CH$_4$.  The steps are in increments of $2\times10^{-6}$ on the D/H ratio.}
\label{ch3d}
\end{center}
\end{figure}

Fig. \ref{ch3d} shows the effect on the variation of the D/H ratio in CH$_4$ on the AKARI spectrum.  The $\nu_6$ vibrational band of CH$_3$D at 1161 cm$^{-1}$ is seen entirely in emission, affecting the shape of the spectrum on the short-wavelength side of the SG2 spectrum.  Unfortunately, the absolute calibration (both in terms of flux, and the wavelength calibration) has the largest degree of uncertainty at the end of the channel:  spectral reductions optimized for the two prominent mid-IR features, C$_2$H$_6$ and CH$_4$, provided differing estimates of the flux in this region, so the optimization for CH$_4$ had to be used.  Furthermore, as no spectral lines are resolved, this can provide no more than an order of magnitude estimate of the CH$_3$D abundance.  Nevertheless, we scaled the CH$_4$ profile by a CH$_3$D/CH$_4$ ratio between $2.0\times10^{-4}$ and $6.0\times10^{-4}$ in steps of $0.1\times10^{-4}$, using the cool, warm and best-fitting $T(p)$ structures from the previous section.  

Minimising $\chi^2$ over the 1100-1200 cm$^{-1}$ spectral range, we find that D/H ratios in CH$_4$  in the range $(3.0\pm1.0)\times10^{-4}$ can fit the spectrum, consistent with the value of $3.25\times10^{-4}$ that we would expect from the ISO-derived D/H ratio in H$_2$ of $6.5\times10^{-5}$ \citep{99feuchtgruber}, using a fractionation factor of $f=1.25$ between deuterium in H$_2$ and CH$_4$ \citep{88fegley}.  This D/H ratio in CH$_4$ is pertinent to the 0.2 mbar level, the peak of the CH$_3$D contribution function in Fig. \ref{cf}

\subsection{Hydrocarbon Emission Features}

\begin{figure}[htb]
\begin{center}
\epsfig{file=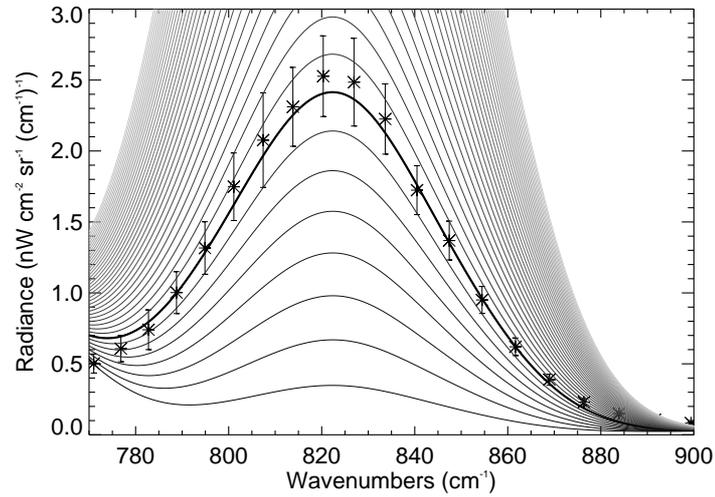,width=10cm}
\caption{Plotting the variation of the modelled radiance compared to the AKARI data for the C$_2$H$_6$ emission feature.  The thick solid line is the model with the best-fit $T(p)$ from Fig. \ref{temp} and 0.8 times the modelled abundance of \citep{05moses_jup}.}
\label{c2h6}
\end{center}
\end{figure}

The second most prominent feature of the mid-IR AKARI spectrum is the ethane peak at 12.2 $\mu$m, which is fitted in Fig. \ref{c2h6}.  Using the three temperature profiles, we scale the C$_2$H$_6$ profile in steps of 0.1 between 0.0 and 2.0, to derive a best fit of $0.8\pm0.2$ times the \textit{a priori} profile.   The error range was calculated using the three models representing the centre and the two extrema of the temperature profiles, yielding 0.6 for the warmest profile, 1.0 for the coldest profile, and 0.8 as the optimum profile.  This yields a maximum abundance of $(2.5\pm0.6)\times10^{-6}$ at 3.4 $\mu$bar, and $(8.5\pm2.1)\times10^{-7}$ at the peak of the ethane contribution function at 0.3 mbar in Fig. \ref{cf}.  

The long-wavelength end of the SG2 spectrum is sensitive to the abundance of acetylene (C$_2$H$_2$), where we see some discrepancy between the model and the data in Fig. \ref{c2h6}.  \citet{05moses_jup} noted the difficulties associated with simultaneously modelling the abundances of C$_2$H$_6$ and C$_2$H$_2$, suggesting that the model overpredicted the abundance of acetylene.  Indeed, a comparison of the model with data in the 760-780 cm$^{-1}$ range implies the need for a scaling factor of 0.5-0.6 to reproduce the observations, suggesting a C$_2$H$_2$ mole fraction of approximately $1.5\times10^{-7}$ at 10 $\mu$bar, or a peak abundance of $2.2\times10^{-7}$ at 3.4 $\mu$bar, though we stress that our sensitivity to C$_2$H$_2$ is limited to the wings of the emission band at 760-790 cm$^{-1}$.

\begin{figure}[htb]
\begin{center}
\epsfig{file=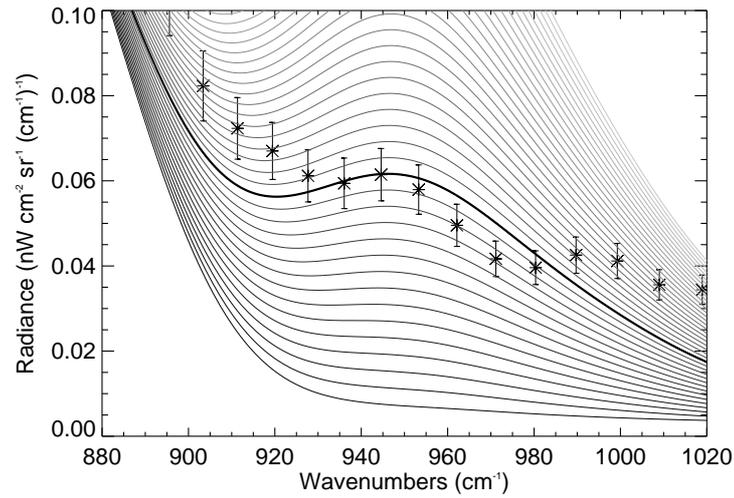,width=10cm}
\caption{Plotting the variation of the modelled radiance compared to the AKARI data for the C$_2$H$_4$ emission feature at 10.5 $\mu$m.  The signal was extremely small between the much larger ethane and methane emission features, but the thick solid line is the model which best-fits the morphology of the spectrum, scaling the model of \citet{05moses_jup} by a factor of 1.4.}
\label{c2h4}
\end{center}
\end{figure}

Finally, the AKARI spectrum between 900-1100 cm$^{-1}$ has a low signal-to-noise ratio (Fig. \ref{spectrum}), because the radiance is very weak at these wavenumbers, but shows a modulation at 952 cm$^{-1}$ (10.5 $\mu$m, the $\nu_7$ band) which could be caused by ethylene (C$_2$H$_4$).  Fig. \ref{c2h4} indicates that the forward model allows us to partially reproduce the morphology of this region of the spectrum.  Using the three temperature profiles, we determine a scale factor for the \citet{05moses_jup} model profile of $1.4^{+0.5}_{-0.6}$, equivalent to a peak abundance of $7.8_{-3.4}^{+2.8}\times10^{-7}$ at 1.3 $\mu$bar, or $5.0_{-2.1}^{+1.8}\times10^{-7}$ at 2.8 $\mu$bar, the peak of the contribution function in Fig. \ref{cf}.  Though we must be cautious, because the spectral morphology is not precisely reproduced by the forward models.  Furthermore, the ISO spectrum presented by \citet{99schulz} showed the C$_2$H$_4$ peak as a narrow single-pixel feature, whereas our modelled ethylene emission for AKARI is rather broad.   Finally, the region of the spectrum between 980-1100 cm$^{-1}$ shows excess emission which is not explained by our model, and it may be possible that higher-order hydrocarbons, whose spectral features are unresolved, are causing additional emission in this range.


\section{Analysis of Near-IR Reflectance Spectra:  1.7-5.5 $\mu$m}
\label{nearir}

Neptune's near-infrared spectrum in Fig. \ref{spectrum} is dominated by reflected sunlight from Neptune's cloud layers, in addition to absorption due to CH$_4$.  The peak at 4.7 $\mu$m is of particular significance, as discussed in Section \ref{CO}.  At shorter wavelengths, the AKARI near-IR spectrum shows considerable deviations from VLT/ISAAC spectra between 3.5-4.1 $\mu$m obtained in August 2002 \citep{03feuchtgruber}, suggestive of meteorological variability discussed in Section \ref{meteo}.  

\subsection{CO Emission}
\label{CO}

\begin{figure}[htb]
\begin{center}
\epsfig{file=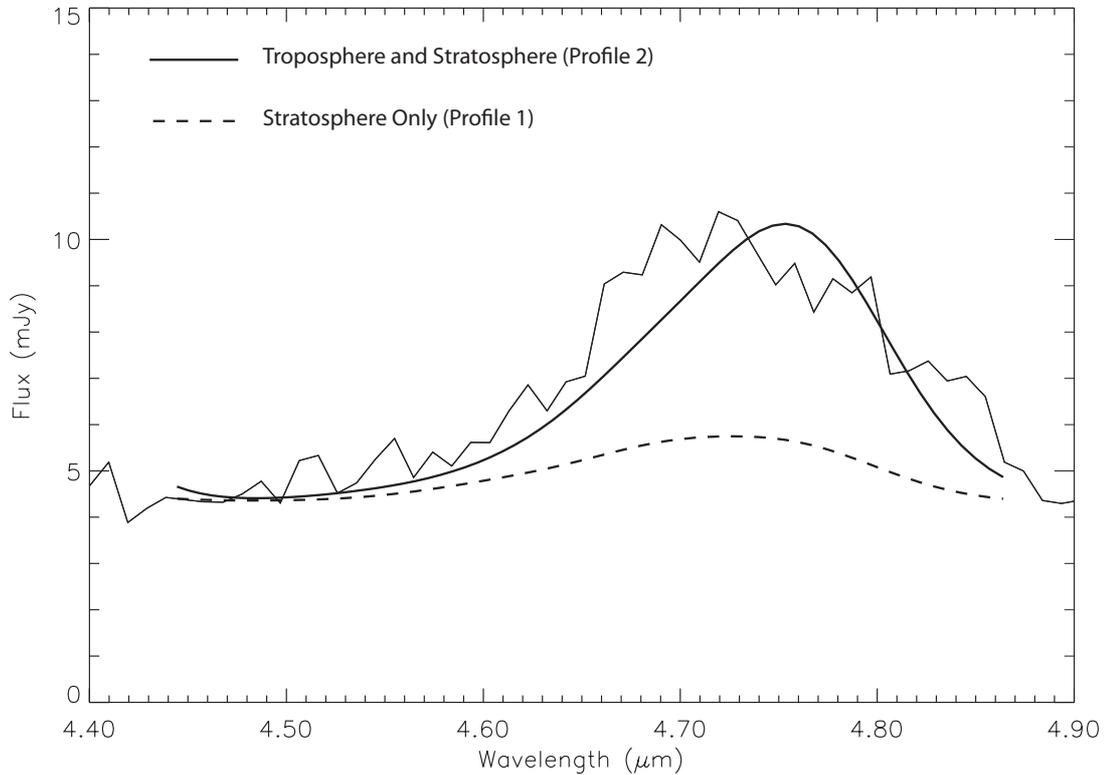,width=15cm}
\caption{Fluorescence of CO compared to the near-IR prism spectrum from AKARI.  The thick solid line is the best fitting profile, with CO present in both the troposphere and stratosphere (profile 2).  The dashed line is the modeled emission with stratospheric CO only, indicating the requirement for tropospheric CO.}
\label{akari_CO}
\end{center}
\end{figure}

The emission peak at 4.7 $\mu$m cannot be reproduced by thermal or solar reflected LTE (local thermodynamic equilibrium) radiative transfer. Just as on Uranus, where a similar emission feature has been detected previously \citep{04encrenaz}, the feature can be interpreted as fluorescent (non-LTE) emission of CO.  The population in the upper levels of the observed transition is by solar absorption, mainly from the CO 1-0 and 2-0 bands.  However, the contribution from the (1-0) band (resonant fluorescence) is strongly self-absorbed, so the (2-1) band therefore dominates the spectrum.  This explains the wavelength shift in the observed spectrum compared to the (1-0) band centre.  A full non-LTE radiative model has  been adapted to Neptune with the following characteristics. The atmospheric structure was taken from the mid-IR determinations in Section \ref{midir}, with a variable CO abundance vertical profile. Two CO vertical profiles were modeled to produce the synthetic spectra in Fig. \ref{akari_CO}:
\begin{itemize}
\item Profile 1:  CO limited to the stratosphere alone with a variable mixing ratio (constant with altitude).   mixing ratio of CO/H$_2$=$2.5\times10^{-6}$, decreasing to zero below 10 mbar.
\item Profile 2:  Similar to the CO profile retrieved from the sub-millimeter range \citep{07hesman}, CO is present in both the stratosphere and troposphere.
\end{itemize}
The upper atmospheric abundance was modeled using an eddy diffusion coefficient of $2\times10^7$ cm$^{2}$s$^{-1}$ \citep{92atreya}.  The synthetic spectra in Fig. \ref{akari_CO} were calculated from a non-LTE radiative model including solar radiation absorption, self-absorption in the resonant fluorescent (1-0) band, and frequency redistribution from vibrational CO bands.  For both profiles, we scaled the abundance of CO to obtain the best fit to the AKARI observations.

The weighting function of the 2-1 fluorescent emission is found to peak around 1 bar, with much smaller contributions from the 1-0 band at higher altitudes (0.01-0.1 mbar), therefore giving sensitivity to the CO distribution in both the stratospheric and tropospheric parts of the atmosphere. A similar depth of emission was found for fluorescence of CO on Uranus, as discussed in \citet{04encrenaz}. Despite the fact that non-LTE effects are usually dominant only in the low-frequency collisional regime of the upper atmosphere, the decrease in the non-LTE contribution is counter-balanced by the increase in the optical depth of CO, and the non-LTE emission in the 2-1 band is not affected by self-absorption like the 1-0 band. 

The results of the non-LTE radiative transfer model are shown in Fig. \ref{akari_CO}.  A continuum has been added to the synthetic spectra to fit the AKARI spectrum outside the CO band.   The dashed line shows the effects of restricting CO to the stratosphere - its absence from the troposphere means that the fluorescent emission is insufficient to reproduce the peak in the AKARI spectrum.  The best-fitting model had CO/H$_2$=$2.5\times10^{-6}$ in the stratosphere with a decrease in the abundance by a factor of four below 10 mbar.  Only by including tropospheric CO do we begin to reproduce the emission feature, which would likely be improved via the addition of higher vibrations CO bands.  However, given the spectral resolution of the observed spectra (and the uncertainties in the input atmospheric profiles discussed earlier), this was not deemed necessary at this stage.  The consequences of tropospheric CO are briefly discussed in Section \ref{discuss}.


\section{Discussion}
\label{discuss}

\subsection{Globally-Averaged Temperatures}

Neptune's stratospheric temperature structure and composition have been previously investigated by a number of authors, and in this section we show that the AKARI SG1 and SG2 modules provide an independent verification of many of these results, particularly of the photochemical models of \citet{05moses_jup}.  Fig. \ref{temp} compares the derived stratospheric temperature to previous studies, and demonstrates convergence of multiple models in the 0.1-100 $\mu$bar range.  The resulting temperatures are at the lower end of the 170-190 K temperature range derived from the Voyager Ultraviolet Spectrometer between 0.3-50 $\mu$bar \citep[UVS,][]{93yelle}.  For pressures greater than 1 mbar the Voyager/RSS results stand out as being cooler  \citep[minimum of 49 K,][]{90lindal_nep} than the others, whereas subsequent analyses of remote sensing data from ISO \citep{98bezard, 03fouchet, 03burgdorf_nep} and the Voyager Infrared Radiometer and Spectrometer \citep[IRIS,][]{98conrath} have suggested warmer tropopause temperatures of approximately 56 K, dropping to 50-52 K at mid-latitudes.  

At pressures less than 1 mbar the AKARI quasi-isothermal temperature profile is remarkably consistent with that used in the photochemical analysis of \citet{05moses_jup}, and warmer in the 0.1-1.0 mbar region than the profile of \citet{98bezard}.  However, the AKARI profile is fully consistent with the ISO/SWS-derived temperature of $170\pm3$ K at 3 $\mu$bar \citep{98bezard}.   The SG1 spectrum also requires that the transition from the quasi-isotherm to the warm temperatures at 0.1 $\mu$bar must occur at pressures lower than 5.0 $\mu$bar so that they do not adversely affect the quality of the spectral fits. The lack of sensitivity at these high altitudes is reflected in the large error bars above 1 $\mu$bar.  The close correspondence of all of the stratospheric temperature models suggests little variability in Neptune's global temperature structure in the decade between the ISO (1997) and AKARI (2007) observations.  But disk-averaged spectroscopy is insensitive to spatial variations in stratospheric temperatures across the planet \citep[e.g. the localised south polar stratospheric emission discussed by][]{07orton_nep}, and it is likely that the compositional and thermal results derived here are weighted towards the high-temperature south polar region.


\subsection{Atmospheric Composition}

Given the large anticorrelation between stratospheric temperatures and the CH$_4$ abundance, it is perhaps surprising that the temperatures match so closely when the methane mole fraction was varied as a free parameter.  With saturation at the coolest tropopause temperatures of 56 K, we would expect a stratospheric mole fraction of $2.2\times10^{-4}$.  The derived CH$_4$ abundance of $(9.0\pm3.0)\times10^{-4}$ at 50 mbar is four times as large as that expected from the cold-trap temperature.  This result is consistent with the modelling of visible CH$_4$ absorption spectra by \citep{94baines}, who found stratospheric abundance in the range $0.2-17.0\times10^{-4}$, with an optimum value of $3.5\times10^{-4}$.  Furthermore, the ranges of CH$_4$ mole fractions quoted by \citet{92orton} ($1.9-26.1\times10^{-4}$) and \citet{07orton_nep} ($7.5-15.0\times10^{-4}$) encompass the AKARI results at mbar pressures.  Most importantly, the optimum value for CH$_4$ is consistent within the errors with the stratospheric abundance of $7.5\times10^{-4}$ used in the photochemical modeling of \citet{05moses_jup}, discussed below.  

\begin{figure}[htb]
\begin{center}
\epsfig{file=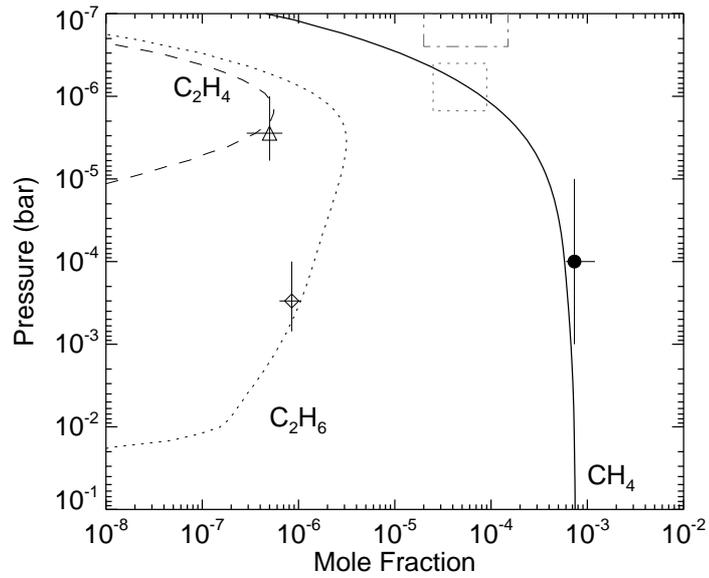,width=10cm}
\caption{Comparison of CH$_4$ (circle, compared to the solid line), C$_2$H$_6$ (diamond, compared to the dotted line) and C$_2$H$_4$ (triangle, compared to the dashed line) results with the predictions of Model A from \citet{05moses_jup}.  The model predictions for the vertical distributions, which were shown to be consistent with Voyager/UVS results in Fig. 32 of \citet{05moses_jup}, were scaled to fit the AKARI spectra.  The dotted box shows the range of CH$_4$ mole fractions from \citet{93yelle}, the dashed box shows the range of values from \citet{92bishop}.}
\label{hydrocarbons}
\end{center}
\end{figure}

Fig. \ref{hydrocarbons} compares the best-fitting abundances of CH$_4$, C$_2$H$_6$ and C$_2$H$_4$ to the predicted vertical profiles of the 1D photochemical model of \citep[][Fig. 32]{05moses_jup}.  The retrieved values have been placed at the peaks of the contribution functions.   If we extrapolate the AKARI-derived CH$_4$ mole fraction to lower pressures using the altitude-dependence of \citet{05moses_jup}, we find values of  $(0.9\pm0.3)\times10^{-4}$ at 1 $\mu$bar.   This is at the upper limit of the $0.25-0.90\times10^{-4}$ range of mole fractions at 0.4-1.5 $\mu$bar derived by \citet{93yelle} from Voyager/UVS data \citep[values extracted from][and shown as a dotted box in Fig. \ref{hydrocarbons}]{05moses_jup}.  Using a similar dataset, \citet{92bishop} found a $0.2-1.5\times10^{-4}$ range between 0.06-0.25 $\mu$bar, which is larger than the estimates from the AKARI-derived values with the model vertical profile (dashed box in Fig. \ref{hydrocarbons}).  AKARI results are more consistent with the \citet{93yelle} study, which improved on the analysis of \citet{92bishop} by using the full 126-166 nm range of the Voyager/UVS data.  So despite the non-uniqueness of the solution and the anti-correlation between the temperature and CH$_4$ abundance, the AKARI thermal-IR results appear to be consistent with the UV-derived results at higher altitudes, confirming the stratospheric super-saturation of CH$_4$.  This suggests that some form of vertical transport of CH$_4$-laden tropospheric air is required to supply methane into the lower stratosphere, as had been previously postulated \citep[e.g.,][]{89lunine}.  This could be the result of vigorous convective activity on a global scale, but it is interesting to note Neptune's south pole is warm enough \citep[62-66 K at 100 mbar from][compared with our nominal value of 56 K]{07orton_nep} to permit CH$_4$ mole fractions of $80-100\times10^{-4}$ to be present in the south polar region, almost an order of magnitude larger than the global mean derived here.  The warm south pole, along with elevated stratospheric CH$_4$, could dominate the disc-averaged spectrum.  CH$_4$ could be subsequently redistributed to other latitudes by stratospheric transport, but spatially-resolved spectroscopy of CH$_4$ is required to distinguish between the effects of localized convection versus polar enhancements.  Disc-averaged AKARI observations are insufficient to distinguish between these two possible mechanisms for the stratospheric enrichment in methane.


We derive an estimate of the D/H ratio in CH$_4$ of $(3.0\pm1.0)\times10^{-4}$ which is consistent with the estimate of $(3.6\pm0.5)\times10^{-4}$ from ground-based spectroscopic observations of \citet{92orton}).  The lower accuracy in our value reflects the uncertainties in the temperature structure and CH$_4$ abundance, along with the absolute flux uncertainties at the short-wavelength end of the AKARI SG2 channel.  Nevertheless, this result is consistent with the value of $3.25^{+1.25}_{-0.75}\times10^{-4}$ that we would expect from the ISO-derived D/H enrichment in H$_2$ of $6.5^{+2.5}_{-1.5}\times10^{-5}$ \citep{99feuchtgruber}, using a fractionation factor of $f=1.25$ between deuterium in H$_2$ and CH$_4$ \citep{88fegley}.  The AKARI analysis confirms the enrichment in Neptune's deuterium abundance relative to the other giant planets.


The ethane mole fraction of $(8.5\pm2.1)\times10^{-7}$ at the peak of the ethane contribution function at 0.3 mbar is smaller than the model predictions by approximately 20\% (Fig. \ref{hydrocarbons}).  Nevertheless, it is consistent with other infrared spectroscopic measurements of the C$_2$H$_6$ emission from heterodyne techniques \citep[$1.9\times10^{-6}$ with a factor of four uncertainty,][]{92kostiuk}, echelle spectroscopy \citep[$1.0^{+0.2}_{-0.8}\times10^{-6}$,][]{92orton} and Voyager/IRIS \citep[$1.5^{+2.5}_{-0.5}\times10^{-6}$, ][]{91bezard}.  Furthermore, the close correspondence between the photochemical model and the AKARI result in Fig. \ref{hydrocarbons} indicates the consistency between thermal-IR and UV results.  \citet{93yelle} used Voyager/UVS to show that the ethane abundance increased from approximately $1\times10^{-6}$ at 1 mbar to $3\times10^{-6}$ at 1 $\mu$bar, which was subsequently used by \citet{05moses_jup} as a constraint on the model curves for Fig. \ref{hydrocarbons}.  Finally, this vertical profile was scaled in the AKARI analysis to obtain the C$_2$H$_6$ mole fraction presented here. 

Because of the low spectral resolution of the AKARI spectra in the 10-$\mu$m region, the abundance of ethylene at 2.8 $\mu$bar (Fig. \ref{hydrocarbons}) should be considered with caution.  Our ethylene estimate of $5.0_{-2.1}^{+1.8}\times10^{-7}$ at 2.8 $\mu$bar is approximately 40\% larger than the model prediction.  Extrapolating to deeper pressures using the vertical dependence of \citet{05moses_jup}, we estimate the 200-$\mu$bar mole fraction to be $1.4^{+0.5}_{-0.6}\times10^{-9}$.   This value is at the upper limit of the range of the C$_2$H$_4$ models tested by \citet{99schulz} in their fitting of ISO photometry data for the same 10-$\mu$m region.  Indeed, \citet{05moses_jup} point out that their photochemical model is more consistent with the UV results of \citet{93yelle}, which showed an increase from $\approx1.0\times10^{-9}$ to $\approx5.0\times10^{-9}$ between 100 $\mu$bar and 10 $\mu$bar.  Given that this vertical profile was scaled by $1.4^{+0.5}_{-0.6}$ to fit the AKARI spectra, we again find good agreement with the photochemical model and UV-derived results.  

An important test of photochemical models is their ability to reproduce the observed ratios of the hydrocarbon species.  The ethane/acetylene, ethane/ethylene and acetylene/ethylene ratios should increase with distance from the Sun because of changing photolysis rates, but a combination of the AKARI results suggest that Model A of \citet{05moses_jup} overestimates the ratios by a factor of 2-3 at 0.2 mbar:  ethane/acetylene is $\approx20$ (model predicts 60), ethane/ethylene is $\approx680$ (model predicts 1180) and acetylene/ethylene is $\approx30$ (model predicts 70).  These difficulties encountered in reproducing these ratios in the photochemical models has been discussed by \citet{05moses_jup}, and it should be borne in mind that the estimates of hydrocarbon abundances depend strongly on the temperature structure, which is itself non-unique.   High spectral resolution observations of the hydrocarbon emission bands may help to resolve these open questions.

At shorter wavelengths in the near-IR, the non-LTE modelling of the emission peak at 4.7 $\mu$m (Fig. \ref{akari_CO}) demonstrated the need for the presence of CO in Neptune's troposphere, independently confirming the interpretation of sub-millimeter observations by \citet{07hesman}.  Although the existence of CO in Neptune's stratosphere has been known since the first millimeter wavelength observations of \citet{93marten}, AKARI is the first instrument to observe the fluorescent emission of CO at 4.7 $\mu$m.  Furthermore, the non-LTE modeling confirms the findings of \citet{07hesman} that an additional reservoir of CO is needed in the upper troposphere to fit the spectra.  Thus Neptune's CO abundance is a combination of CO from the thermochemistry of carbon and oxygen in the deep atmosphere, advected to altitudes accessible to remote sensing, in addition to stratospheric CO from external sources.  Our CO distribution (CO/H$_2$=$2.5\times10^{-6}$ in the stratosphere, decreasing by a factor of four below 10 mbar) is entirely consistent with the best-fitting result of \citet{07hesman}.  The fluorescent emission spectra are inconsistent with the constant vertical CO distribution proposed by \citet{93marten} to fit the first measurements of CO emission in the millimeter range.  However, as noted by \citet{07hesman}, these early results were sensitive only to the emission cores, and not to tropospheric absorption further from the line centre, so that the constant vertical profile was a perfectly adequate assumption.  The AKARI modeling confirms that such an assumption is no longer valid.

CO is a disequilibrium species in the upper troposphere (it is not thermochemically stable at the temperatures and pressures observed), so must be supplied by convective mixing from the deeper atmosphere.  On the other hand, stratospheric CO is probably the result of infalling exogenic material, either supplying CO directly \citep[a cometary impact hypothesis, ][]{05lellouch} or by supplying H$_2$O to react with CH$_4$ photochemical products \citep[meteoritic bombardment,][]{92rosenqvist}.  As suggested by \citet{07hesman}, the determination of the CO profile alone is insufficient for discrimination between the different external sources.  Spatially resolved spectroscopy of CO and other O-bearing species would provide invaluable clues to understand the stratospheric CO abundance.

\subsection{Meteorological Variations}
\label{meteo}

\citet{03feuchtgruber} measured the spectrum of Neptune in August 2002 in the L-band (3.5-4.1 $\mu$m) with the ISAAC imaging spectrometer at the ANTU of the Very Large Telescope at the European Southern Observatory in Cerro Paranal.   They found the observed flux to be stronger by a factor of three compared to measurements at the same wavelength recorded in 1997, five years earlier, and attributed the excess of infrared flux to an increased production of photochemical aerosols. This result suggested that Neptune's meteorology was actually linked to the variation of the solar incident flux.   A similar conclusion was drawn from visible photometry of Neptune, which is correlated with the solar activity cycle via solar UV as a causative mechanism \citep{02lockwood}.   In addition to these long-term trends, however, there were also isolated infrared outbursts, for example in 1976 \citep{77pilcher}. In contrast, the AKARI measurements in the same spectral range produces flux densities 4 times \textbf{smaller} than those measured 5 years earlier by \citet{03feuchtgruber}.

In Table \ref{SSN} we compare the average flux density of Neptune in the L-band as measured by AKARI with the previous findings (estimated from spectra in the relevant publications), and with the number of sunspots (representative of the solar variability for the incident flux).  The correlation coefficient between flux density and sunspot number for the latest three dates is 0.99. This is surprisingly high given the considerable uncertainties in both variables, but supports the connection between L-band flux density (related to the generation of photochemical aerosols in the stratosphere) and solar incident flux discussed by \citet{03feuchtgruber}.  


\section{Conclusions}

Infrared spectra from the AKARI Infrared Camera between 1.8 and 13 $\mu$m have been analysed to determine the stratospheric temperature structure and composition.  The 7.7-$\mu$m CH$_4$ emission band was used to simultaneously derive a quasi-isothermal temperature structure (consistent with previous studies in the 1-1000 $\mu$bar range) and a CH$_4$ stratospheric mole fraction of $(9.0\pm3.0)\times10^{-4}$ at 50 mbar, decreasing to $(0.9\pm0.3)\times10^{-4}$ at 1 $\mu$bar.  The close agreement in the globally averaged temperature structure derived from ISO-observations in 1997 and AKARI-observations in 2007 suggest negligible changes to Neptune's global stratospheric temperature structure in the intervening decade.  The disk-averaged nature of the AKARI spectra prevent detection of spatial variations in the temperature field, though we expect these conclusions to be weighted towards the warm south polar region of Neptune where the hydrocarbon emissions are strongest \citep{07orton_nep}.  The CH$_4$ abundance is larger than that expected if the tropopause at 56 K is efficiently cold-trapping methane in the troposphere, but consistent with the hypothesis that CH$_4$ leaking through the warm south polar tropopause \citep[62-66 K, ][]{07orton_nep} is globally redistributed by stratospheric motion.  

The uncertainties on the retrieved $T(p)$ profile are large because of the strong anticorrelation of temperature and CH$_4$ abundance.  However, this was subsequently used to derive (a) the ratio of D/H in CH$_4$ of $3.0\pm1.0\times10^{-4}$ from the CH$_3$D emission at 8.6 $\mu$m; (b) the stratospheric mole fraction of ethane of $(8.5\pm2.1)\times10^{-7}$ at 0.3 mbar from 12.2-$\mu$m; and (c) the mole fraction of ethylene of $5.0_{-2.1}^{+1.8}\times10^{-7}$ at 2.8 $\mu$bar from the 10.5-$\mu$m emission band.  The D/H ratio supports the conclusion that Neptune was enriched in deuterium relative to the other gas giants at the time of the planets formation \citep{99feuchtgruber}.  The observed ratios of the hydrocarbon mole fractions will provide important inputs to constrain photochemical modeling for Neptune's stratosphere, and explain the observed differences between the giant planets.

At shorter wavelengths, Neptune flux densities in the 3.5-4.1 $\mu$m range \citep[indicative of the production of stratospheric hydrocarbon hazes, ][]{03feuchtgruber} show considerable variation with time.  Comparison with previous studies confirms that the flux density is highly correlated with solar variability (determined from the solar sunspot cycle), with the low flux measured by AKARI the direct result of the present solar minimum.  AKARI spectra demonstrate the first evidence for Neptune's fluorescent emission of CO at 4.7 $\mu$m, comparable with similar emission observed recently on Uranus \citep{04encrenaz}.  The fluorescent emission can only be reproduced using a vertical CO distribution similar to that derived from sub-millimeter wavelengths by \citet{07hesman}, confirming the necessity for two sources of CO in Neptune's atmosphere, one from convective mixing with the deep troposphere, and another exogenic source for the stratosphere.

\begin{acknowledgements}
This work is based on observations with AKARI, a JAXA project with the participation of ESA. Fletcher was supported by an appointment to the NASA Postdoctoral Program at the Jet Propulsion Laboratory, administered by Oak Ridge Associated Universities through a contract with NASA.  Orton carried out part of this research at the Jet Propulsion Laboratory, California Institute of Technology, under a contract with NASA. 
\end{acknowledgements}

\bibliographystyle{aa}

\bibliography{13358.ref}

\begin{thebibliography}{44}
\expandafter\ifx\csname natexlab\endcsname\relax\def\natexlab#1{#1}\fi

\bibitem[{{Atreya}(1992)}]{92atreya}
{Atreya}, S.~K. 1992, Advances in Space Research, 12, 11

\bibitem[{Baines \& Hammel(1994)}]{94baines}
Baines, K. \& Hammel, H. 1994, Icarus, 109, 20

\bibitem[{{Baines} {et~al.}(1995){Baines}, {Mickelson}, {Larson}, \&
  {Ferguson}}]{95baines}
{Baines}, K.~H., {Mickelson}, M.~E., {Larson}, L.~E., \& {Ferguson}, D.~W.
  1995, Icarus, 114, 328

\bibitem[{{B{\'e}zard} {et~al.}(1998){B{\'e}zard}, {Feuchtgruber}, {Moses}, \&
  {Encrenaz}}]{98bezard}
{B{\'e}zard}, B., {Feuchtgruber}, H., {Moses}, J.~I., \& {Encrenaz}, T. 1998,
  Astron. Astrophys., 334, L41

\bibitem[{{B{\'e}zard} {et~al.}(1991){B{\'e}zard}, {Romani}, {Conrath}, \&
  {Maguire}}]{91bezard}
{B{\'e}zard}, B., {Romani}, P.~N., {Conrath}, B.~J., \& {Maguire}, W.~C. 1991,
  \jgr, 96, 18961

\bibitem[{{Bishop} {et~al.}(1992){Bishop}, {Atreya}, {Romani}, {Sandel}, \&
  {Herbert}}]{92bishop}
{Bishop}, J., {Atreya}, S.~K., {Romani}, P.~N., {Sandel}, B.~R., \& {Herbert},
  F. 1992, \jgr, 97, 11681

\bibitem[{Burgdorf {et~al.}(2003)Burgdorf, Orton, Davis, Sidher, Feuchtgruber,
  Griffin, \& Swinyard}]{03burgdorf_nep}
Burgdorf, M., Orton, G., Davis, G., {et~al.} 2003, Icarus, 164, 244

\bibitem[{Conrath {et~al.}(1993)Conrath, Gautier, Owen, \&
  Samuelson}]{93conrath_nep}
Conrath, B., Gautier, D., Owen, T., \& Samuelson, R. 1993, Icarus, 101, 168

\bibitem[{{Conrath} {et~al.}(1998){Conrath}, {Gierasch}, \&
  {Ustinov}}]{98conrath}
{Conrath}, B.~J., {Gierasch}, P.~J., \& {Ustinov}, E.~A. 1998, Icarus, 135, 501

\bibitem[{{Encrenaz} {et~al.}(2004){Encrenaz}, {Lellouch}, {Drossart},
  {Feuchtgruber}, {Orton}, \& {Atreya}}]{04encrenaz}
{Encrenaz}, T., {Lellouch}, E., {Drossart}, P., {et~al.} 2004, \aap, 413, L5

\bibitem[{{Encrenaz} {et~al.}(2000{\natexlab{a}}){Encrenaz}, {Schulz},
  {Drossart}, {Lellouch}, {Feuchtgruber}, \& {Atreya}}]{00encrenaz}
{Encrenaz}, T., {Schulz}, B., {Drossart}, P., {et~al.} 2000{\natexlab{a}},
  \aap, 358, L83

\bibitem[{{Encrenaz} {et~al.}(2000{\natexlab{b}}){Encrenaz}, {Schulz},
  {Drossart}, {Lellouch}, {Feuchtgruber}, \& {Atreya}}]{00encrenaz_esa}
{Encrenaz}, T., {Schulz}, B., {Drossart}, P., {et~al.} 2000{\natexlab{b}}, in
  ESA Special Publication, Vol. 456, ISO Beyond the Peaks: The 2nd ISO Workshop
  on Analytical Spectroscopy, ed. A.~{Salama}, M.~F. {Kessler}, K.~{Leech}, \&
  B.~{Schulz}, 19--+

\bibitem[{{Fegley} \& {Prinn}(1988)}]{88fegley}
{Fegley}, B. \& {Prinn}, R.~G. 1988, Astrophys. J., 326, 490

\bibitem[{{Feuchtgruber} \& {Encrenaz}(2003)}]{03feuchtgruber}
{Feuchtgruber}, H. \& {Encrenaz}, T. 2003, A\&A, 403, L7

\bibitem[{{Feuchtgruber} {et~al.}(1999){Feuchtgruber}, {Lellouch},
  {B{\'e}zard}, {Encrenaz}, {de Graauw}, \& {Davis}}]{99feuchtgruber}
{Feuchtgruber}, H., {Lellouch}, E., {B{\'e}zard}, B., {et~al.} 1999, Astron.
  Astrophys., 341, L17

\bibitem[{{Fletcher} {et~al.}(2009{\natexlab{a}}){Fletcher}, {Orton}, {Teanby},
  \& {Irwin}}]{09fletcher_ph3}
{Fletcher}, L.~N., {Orton}, G.~S., {Teanby}, N.~A., \& {Irwin}, P.~G.~J.
  2009{\natexlab{a}}, Icarus, 202, 543

\bibitem[{{Fletcher} {et~al.}(2009{\natexlab{b}}){Fletcher}, {Orton}, {Teanby},
  {Irwin}, \& {Bjoraker}}]{09fletcher_ch4}
{Fletcher}, L.~N., {Orton}, G.~S., {Teanby}, N.~A., {Irwin}, P.~G.~J., \&
  {Bjoraker}, G.~L. 2009{\natexlab{b}}, Icarus, 199, 351

\bibitem[{{Fouchet} {et~al.}(2003){Fouchet}, {Lellouch}, \&
  {Feuchtgruber}}]{03fouchet}
{Fouchet}, T., {Lellouch}, E., \& {Feuchtgruber}, H. 2003, Icarus, 161, 127

\bibitem[{{Goody} \& {Yung}(1989)}]{89goody}
{Goody}, R.~M. \& {Yung}, Y.~L. 1989, {Atmospheric radiation : theoretical
  basis} (Atmospheric radiation : theoretical basis, 2nd ed., by Richard
  M.~Goody and Y.L.~Yung.~ New York, NY: Oxford University Press, 1989)

\bibitem[{Grevesse {et~al.}(2007)Grevesse, Asplund, \& Sauval}]{07grevesse}
Grevesse, N., Asplund, M., \& Sauval, A. 2007, Space Science Reviews, 130, 105

\bibitem[{{Guillot}(1999)}]{99guillot}
{Guillot}, T. 1999, Science, 286, 72

\bibitem[{{Hersant} {et~al.}(2004){Hersant}, {Gautier}, \&
  {Lunine}}]{04hersant}
{Hersant}, F., {Gautier}, D., \& {Lunine}, J.~I. 2004, Plan. \& Space Sci., 52,
  623

\bibitem[{{Hesman} {et~al.}(2007){Hesman}, {Davis}, {Matthews}, \&
  {Orton}}]{07hesman}
{Hesman}, B.~E., {Davis}, G.~R., {Matthews}, H.~E., \& {Orton}, G.~S. 2007,
  Icarus, 186, 342

\bibitem[{{Irwin} {et~al.}(1997){Irwin}, {Calcutt}, \& {Taylor}}]{97irwin}
{Irwin}, P.~J.~G., {Calcutt}, S.~B., \& {Taylor}, F.~W. 1997, Adv. Space Res.,
  19, 1149

\bibitem[{{Kostiuk} {et~al.}(1992){Kostiuk}, {Romani}, {Espenak}, \&
  {Bezard}}]{92kostiuk}
{Kostiuk}, T., {Romani}, P., {Espenak}, F., \& {Bezard}, B. 1992, Icarus, 99,
  353

\bibitem[{{Lellouch} {et~al.}(2005){Lellouch}, {Moreno}, \&
  {Paubert}}]{05lellouch}
{Lellouch}, E., {Moreno}, R., \& {Paubert}, G. 2005, \aap, 430, L37

\bibitem[{Lindal {et~al.}(1990)Lindal, Lyons, Sweemam, Eshleman, Hinson, \&
  Tyler}]{90lindal_nep}
Lindal, G., Lyons, J., Sweemam, D., {et~al.} 1990, Geophys. Res. Lett., 17,
  1733

\bibitem[{{Lockwood} \& {Thompson}(2002)}]{02lockwood}
{Lockwood}, G.~W. \& {Thompson}, D.~T. 2002, Icarus, 156, 37

\bibitem[{{Lunine} \& {Hunten}(1989)}]{89lunine}
{Lunine}, J.~I. \& {Hunten}, D.~M. 1989, \planss, 37, 151

\bibitem[{{Marten} {et~al.}(1993){Marten}, {Gautier}, {Owen}, {Sanders},
  {Matthews}, {Atreya}, {Tilanus}, \& {Deane}}]{93marten}
{Marten}, A., {Gautier}, D., {Owen}, T., {et~al.} 1993, \apj, 406, 285

\bibitem[{{Meadows} {et~al.}(2008){Meadows}, {Orton}, {Line}, {Liang}, {Yung},
  {van Cleve}, \& {Burgdorf}}]{08meadows}
{Meadows}, V.~S., {Orton}, G., {Line}, M., {et~al.} 2008, Icarus, 197, 585

\bibitem[{{Moses} {et~al.}(2005){Moses}, {Fouchet}, {B{\'e}zard}, {Gladstone},
  {Lellouch}, \& {Feuchtgruber}}]{05moses_jup}
{Moses}, J.~I., {Fouchet}, T., {B{\'e}zard}, B., {et~al.} 2005, Journal of
  Geophysical Research (Planets), 110, 8001

\bibitem[{{Murakami} {et~al.}(2007){Murakami}, {Baba}, {Barthel}, {Clements},
  {Cohen}, {Doi}, {Enya}, {Figueredo}, {Fujishiro}, {Fujiwara}, {Fujiwara},
  {Garcia-Lario}, {Goto}, {Hasegawa}, {Hibi}, {Hirao}, {Hiromoto}, {Hong},
  {Imai}, {Ishigaki}, {Ishiguro}, {Ishihara}, {Ita}, {Jeong}, {Jeong},
  {Kaneda}, {Kataza}, {Kawada}, {Kawai}, {Kawamura}, {Kessler}, {Kester},
  {Kii}, {Kim}, {Kim}, {Kobayashi}, {Koo}, {Kwon}, {Lee}, {Lorente}, {Makiuti},
  {Matsuhara}, {Matsumoto}, {Matsuo}, {Matsuura}, {M{\"u}ller}, {Murakami},
  {Nagata}, {Nakagawa}, {Naoi}, {Narita}, {Noda}, {Oh}, {Ohnishi}, {Ohyama},
  {Okada}, {Okuda}, {Oliver}, {Onaka}, {Ootsubo}, {Oyabu}, {Pak}, {Park},
  {Pearson}, {Rowan-Robinson}, {Saito}, {Sakon}, {Salama}, {Sato}, {Savage},
  {Serjeant}, {Shibai}, {Shirahata}, {Sohn}, {Suzuki}, {Takagi}, {Takahashi},
  {Tanab{\'e}}, {Takeuchi}, {Takita}, {Thomson}, {Uemizu}, {Ueno}, {Usui},
  {Verdugo}, {Wada}, {Wang}, {Watabe}, {Watarai}, {White}, {Yamamura},
  {Yamauchi}, \& {Yasuda}}]{07murakami}
{Murakami}, H., {Baba}, H., {Barthel}, P., {et~al.} 2007, \pasj, 59, 369

\bibitem[{{Onaka} {et~al.}(2007){Onaka}, {Matsuhara}, {Wada}, {Fujishiro},
  {Fujiwara}, {Ishigaki}, {Ishihara}, {Ita}, {Kataza}, {Kim}, {Matsumoto},
  {Murakami}, {Ohyama}, {Oyabu}, {Sakon}, {Tanab{\'e}}, {Takagi}, {Uemizu},
  {Ueno}, {Usui}, {Watarai}, {Cohen}, {Enya}, {Ootsubo}, {Pearson}, {Takeyama},
  {Yamamuro}, \& {Ikeda}}]{07onaka}
{Onaka}, T., {Matsuhara}, H., {Wada}, T., {et~al.} 2007, \pasj, 59, 401

\bibitem[{Orton {et~al.}(2007)Orton, Encrenaz, Leyrat, Puetter, \&
  Friedson}]{07orton_nep}
Orton, G., Encrenaz, T., Leyrat, C., Puetter, R., \& Friedson, A. 2007, A\&A,
  473, L5

\bibitem[{{Orton} {et~al.}(1987){Orton}, {Aitken}, {Smith}, {Roche},
  {Caldwell}, \& {Snyder}}]{87orton}
{Orton}, G.~S., {Aitken}, D.~K., {Smith}, C., {et~al.} 1987, Icarus, 70, 1

\bibitem[{{Orton} {et~al.}(1990){Orton}, {Baines}, {Caldwell}, {Romani},
  {Tokunaga}, \& {West}}]{90orton}
{Orton}, G.~S., {Baines}, K.~H., {Caldwell}, J., {et~al.} 1990, Icarus, 85, 257

\bibitem[{{Orton} {et~al.}(1992){Orton}, {Lacy}, {Achtermann}, {Parmar}, \&
  {Blass}}]{92orton}
{Orton}, G.~S., {Lacy}, J.~H., {Achtermann}, J.~M., {Parmar}, P., \& {Blass},
  W.~E. 1992, Icarus, 100, 541

\bibitem[{Owen \& Encrenaz(2006)}]{06owen}
Owen, T. \& Encrenaz, T. 2006, Planetary and Space Science, 54, 1188

\bibitem[{{Pilcher}(1977)}]{77pilcher}
{Pilcher}, C.~B. 1977, \apj, 214, 663

\bibitem[{{Rosenqvist} {et~al.}(1992){Rosenqvist}, {Lellouch}, {Romani},
  {Paubert}, \& {Encrenaz}}]{92rosenqvist}
{Rosenqvist}, J., {Lellouch}, E., {Romani}, P.~N., {Paubert}, G., \&
  {Encrenaz}, T. 1992, \apjl, 392, L99

\bibitem[{{Schulz} {et~al.}(1999){Schulz}, {Encrenaz}, {B{\'e}zard}, {Romani},
  {Lellouch}, \& {Atreya}}]{99schulz}
{Schulz}, B., {Encrenaz}, T., {B{\'e}zard}, B., {et~al.} 1999, \aap, 350, L13

\bibitem[{{Sromovsky} {et~al.}(2001){Sromovsky}, {Fry}, {Dowling}, {Baines}, \&
  {Limaye}}]{01sromovsky}
{Sromovsky}, L.~A., {Fry}, P.~M., {Dowling}, T.~E., {Baines}, K.~H., \&
  {Limaye}, S.~S. 2001, Icarus, 149, 459

\bibitem[{{Yelle} {et~al.}(1993){Yelle}, {Herbert}, {Sandel}, {Vervack}, \&
  {Wentzel}}]{93yelle}
{Yelle}, R.~V., {Herbert}, F., {Sandel}, B.~R., {Vervack}, Jr., R.~J., \&
  {Wentzel}, T.~M. 1993, Icarus, 104, 38

\end{thebibliography}



%

\clearpage
\onecolumn

\begin{table}
\caption{Akari/IRC Neptune Data}             
\label{obs}      
\centering                          
\begin{tabular}{c c c c c}        
\hline\hline                 
Module & UT Time & Subsolar Longitude & Wavelength & Resolution \\    
\hline                        
NG & 1756-1826 & 165 & 2.5-5.0 $\mu$m & 135 at 3.6 $\mu$m\\
NP & 0302-0332 & 189 & 1.7-5.5 $\mu$m & 22 at 3.5 $\mu$m\\
SG1 & 0302-0332 & 189 & 5.5-8.3 $\mu$m & 47 at 6.6 $\mu$m\\
SG2 & 0302-0332 &189 & 7.4-13 $\mu$m & 34 at 10.6 $\mu$m\\

\hline                                   
\end{tabular}
\end{table}

\begin{table}
\caption{Correlation between L-band flux density and sunspot number (SSN).  L-band flux densities were estimated from the spectra shown in the relevant publications [1] \citet{77pilcher}, [2] \citet{00encrenaz}, [3] \citet{03feuchtgruber} and [4] This work.   Monthly averages and standard deviation for SSN were taken from the International Sunspot Numbers, compiled by the Solar Influences Data Analysis Center in Belgium (http://solarscience.msfc.nasa.gov/greenwch/spot\_num.txt).          }   
\label{SSN}      
\centering                          
\begin{tabular}{c c c c}        
\hline\hline                 
Date & L-Band Flux Density & SSN & Reference \\
 & mJy & & \\
\hline                        
  May 1975 (Kitt Peak) & $\leq$~10 & 9~$\pm$~11.1 & [1] \\
  May 1997 (ISOPHOT) & 7 & 18.5~$\pm$~13.4 & [2]\\
  Aug 2002 (VLT) & 20 & 116.4~$\pm$~36.3 & [3]\\
  May 2007 (AKARI) & 5.5 & 11.7~$\pm$~6.8 & [4] \\
\hline                                   
\end{tabular}
\end{table}

\end{document}